# 硅纳米团簇与石墨烯复合结构储锂性能的第一性原理研究[*]

吴江滨，钱耀，郭小杰，崔先慧，缪灵[†]，江建军

(华中科技大学 电子科学与技术系 430074)

本文采用第一性原理计算方法，研究了不同晶向硅纳米团簇与石墨烯复合结构稳定性及其储锂性能。计算了不同高度、大小硅团簇与石墨烯复合结构的结合能，复合结构中嵌锂吸附能和PDOS。分析表明，硅团簇和石墨烯之间形成较强的Si-C键，其中[111]晶向硅团簇与石墨烯作用的形成能最高，结构最为稳定。进一步计算其嵌锂吸附能，发现硅团簇中靠近石墨烯界面处的储锂位置更加有利于锂的吸附，由于锂和碳、硅之间有较强电荷转移，其吸附能明显大于其他储锂位置。同时在锂嵌入过程中，由于石墨烯的引入，明显减小了界面处硅的形变，有望提高了其作为锂电池负极材料的可逆容量。

**关键词**：石墨烯　　硅纳米团簇　　第一性原理　　锂离子电池

**PACC**：6140M， 　6855， 　7115A， 　　8640H

**PACS**：36.40.Cg， 　63.20.dk， 　71.15 .Mb， 　84.60.Ve

## 1. 引言

锂离子电池具有体积小、重量轻、功率密度高等优良特点，如何进一步提高其能量密度、功率密度和循环寿命是当前的研究热点[1]。相比于其他材料[2][3][4][5]，硅作为锂电池负极材料其理论电化学容量可以达到 4200 mAh/g[6]，受到广泛关注。然而，硅材料在锂离子嵌入后会产生很大的体积膨胀和显著的结构应力，严重影响了锂电池循环寿命和能量密度。可以通过制备纳米形态的硅材料，如纳米线[7]、纳米片层[8]、纳米空心球[9]、纳米团簇[10]来减小锂电池充放电时的电极形变，同时显著减少了锂嵌入时的行程，提高充放电速度。Zhang Q F 和 Chan T L 分别通过第一性原理计算，理论上探究了锂在硅纳米线[11]和硅纳米团簇[12]中嵌入和扩散机理。在纳米硅材料中最突出的是 Candace K[7]等提出的硅纳米线，第一次循环容量高达 4277 mAh/g，但是在 20 次循环后只剩 3500 mAh/g，在循环寿命上有很大的提高空间。

碳纳米材料由于其巨大的比表面积，也具有较好的储锂性能。1998 年，Che G L 等[13]最先报道了多壁碳纳米管的储锂研究，其嵌锂容量达到 490 mAh/g，1999 年 Frackowiak E 等[14]也得到了相似的结果。用不同方法制备出来的单壁碳纳米管，未经处理前的储锂容量约为 446 mAh/g [15] [16]。2004 年石墨烯被发现后[17]，研究表明其同样具有较好的储锂性能[18]。而且石墨烯的良好力学特性使得其在充放电过程中形变很小，从而大大提高了锂电池的可逆容量[19] [20][21]。

因此，许多研究将硅和碳纳米材料复合，利用碳纳米材料良好的力学性能来减缓硅在锂嵌入时产生的形变。Cui L F 和 Wang W 分别将硅纳米结构与碳纳米管[22] [23]复合用于锂电池负极，得到了较好的电化学容量和循环寿命。Wang X L[24]小组和 Xiang H F[25]小组分别通过实验制备了硅纳米结构与石墨烯的复合材料，结果均表明其电化学容量可达到 2470 mhA/g 以上，且在循环多次之后有较多容量剩余。

此前关于硅纳米结构与石墨烯复合材料作为负极材料的储锂性能较多实验研究，而对于锂在其中嵌入机理的理论研究较为少见。本文将通过第一性原理计算方法，系统地研究硅纳米团簇与石墨烯复合结构的稳定性，从中选取最为稳定的结构，讨论其储锂的机理，以及锂的嵌入对结构的影响，从而提高锂电池的循环性能。

## 2. 模型与计算方法

本文讨论了不同晶向（[100],[110]和[111]）硅团簇与石墨烯作用的复合结构，如图 1 所示。同时我们也分别研究了硅团簇不同直径（约为 0.5 nm，0.75 nm 和 1 nm）和不同高度（约为 0.4 nm，0.7 nm，1.0 nm 和 1.2 nm）时的情况。周期性结构中，两相邻硅团簇之间距离大于 6 Å，在垂直于石墨烯平面的方向，保证硅团簇顶端到晶格顶端有 8 Å 以上的真空层，以保证消除两者之间的相互作用。在复合结构硅团簇中加入锂原子时有两种位置，如图 1(a)中的四面体中心 T 位和六角形中心 H 位。

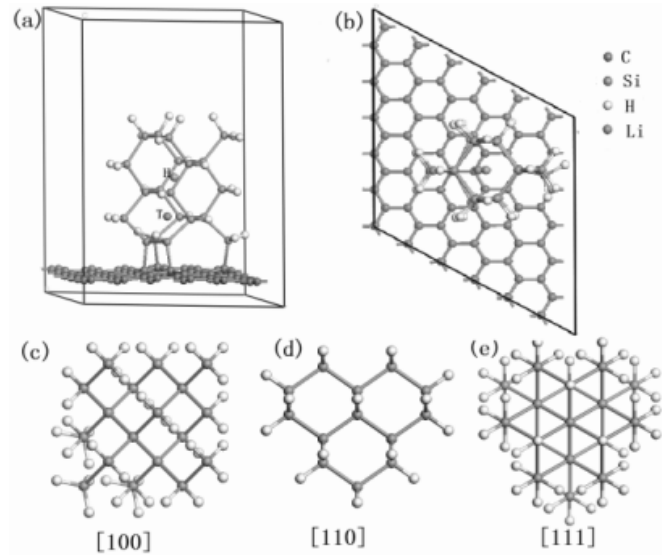

图 1 硅纳米团簇石墨烯复合模型(a)侧视图(b)顶视图。(c)(d)(e)不同硅团簇晶向

本文计算采用基于密度泛函理论[26][27]的 Siesta 软件包[28]，交换关联泛函为局域密度近似[29]中的 CA－PZ[30]近似方法，采用了双 Zeta 极化轨道函数作为基组。基矢展开波函数截断能 $E_c$ 为 150 Ryd。在结构优化时，原子受力的收敛精度为 0.05 eV/ Å。对于硅团簇直径为 0.5 nm 的模型 K 点网格设置为 2×2×1，而直径大于 0.5 nm 的模型 K 点网格设置为 1×1×1。

## 3. 结果与讨论

### 3.1 硅纳米团簇和石墨烯的复合结构

我们研究了不同高度、不同直径和不同晶向的硅纳米团簇与石墨烯的复合结构。硅晶体与石墨烯晶格并不完全匹配，不同晶相匹配程度不同。图 2(a)所示的是直径为 0.75 nm，高度为 1.6 nm[111]晶向的硅团簇与石墨烯作用而形成的结构。结构优化后，界面处硅团簇与石墨烯均发生明显形变。图 2(b)给出了界面处成键 C、Si 原子周围的差分电荷密度图。可见在 C 原子和 Si 原子之间具有很高的电荷密度，表明 Si 原子与 C 原子形成共价键。图 2(c)给出了界面处 Si 原子与 C 原子的 PDOS，可见 C 原子的 2p 轨道和相邻 Si 原子的 3p 轨道有较强交叠，因此推断出 C 和 Si 形成了较强的相互作用。形成的 C-Si 键平均键长为 2.02 Å，比 SiC 中的 C－Si 键（1.891 Å）稍长（6.8%）。

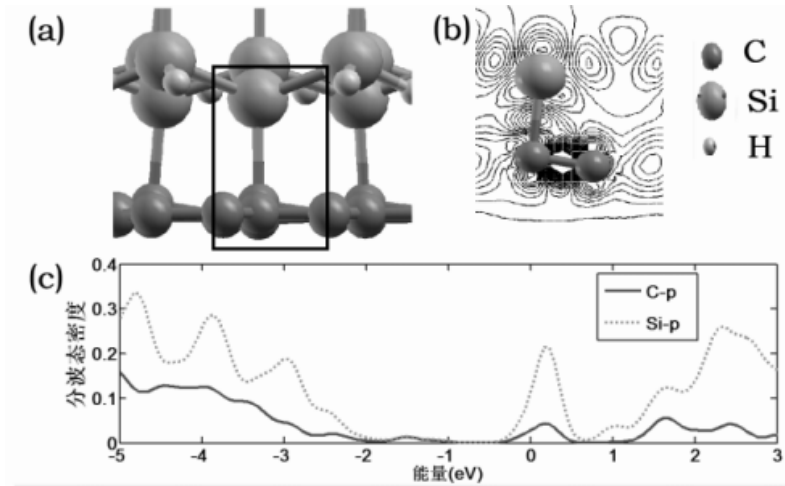

图 2(a)界面处原子成键情况，(b)成键原子周围差分电荷密度，(c)成键 C、Si 原子 PDOS

为进一步衡量复合结构的稳定性，表 1 给出了其结合能 $E = E_0 - E_1 - E_2$。式中，$E$ 为结构的结合能，$E_0$ 为复合结构的总能，$E_1$ 为石墨烯的总能，$E_2$ 为硅团簇的总能。定义单键能 $E_d = E/n$，$n$ 为 C－Si 键的数目。复合结构的结合能随着硅团簇尺寸的增大有着增大的趋势。因为硅团簇越大，与石墨烯成的键越多，结合能越大，稳定性越强。其单键能的范围为－0.71 eV～－1.56 eV，可见硅团簇与墨烯形成了较强的相互作用。相对于[100]和[110]晶向，[111]晶向的单键能较大，所以当[111]晶向与石墨烯作用时，所形成的结构是最为稳定的。这是因为硅的(111)晶面也为六角晶格，和石墨烯作用时原子的相对位置最为匹配，对石墨烯结构破坏最小，容易成键。

表 1 硅纳米团簇—石墨烯复合模型的结合能和单键能（eV）

| 尺寸 | 结合能 | | | 单键能 | | |
| --- | --- | --- | --- | --- | --- | --- |
| | [100] | [110] | [111] | [100] | [110] | [111] |
| 0.5 nm | －2.32 | －3.58 | －4.67 | －0.77 | －0.60 | －1.56 |
| 0.75 nm | －2.94 | －2.70 | －4.55 | －0.73 | －0.54 | －1.52 |
| 1.0 nm | －4.99 | －5.71 | －8.96 | －0.71 | －0.48 | －1.49 |

我们进一步研究了在同一种复合结构上成键数量不同的情况。对于每一晶向，随着团簇尺寸增大，成键数目增多，体系的稳定性增强，同时单键能减小。

表 2 给出了硅团簇直径为 1.0 nm，[111]晶向时的复合结构在成不同数目键时的结合能和单键能。随着成键数目的增加，复合结构的结合能呈现增大的趋势，单键能逐渐的减小。因为当硅团簇与石墨烯成的键越多，释放的能量越大，体系越稳定。而硅团簇与石墨烯成键时破坏了硅团簇和石墨烯的有序性，当成键越多对于这种有序性的破坏就越强烈，单键能越低。

表 2 [111]的硅团簇与石墨烯复合模型的成键数目、结合能和单键能

| 键的数目 | 2 | 3 | 4 | 5 | 6 |
| --- | --- | --- | --- | --- | --- |
| 结合能（eV） | —7.52 | —8.41 | —8.07 | —8.82 | —8.96 |
| 单键能（eV） | —3.76 | —2.80 | —2.02 | —1.76 | —1.49 |

对不同高度硅团簇复合结构的结合能也进行了研究。对于直径为 0.5 nm，作用晶向为[100]的硅团簇，当团簇高度为 0.4 nm、0.7 nm、1.0 nm 和 1.2 nm 时，其结合能分别为－2.267 eV、－2.408 eV、－2.368 eV 和－2.076 eV。可见随着高度的增加结合能并没有显著的变化，复合的结合能对团簇的高度不敏感。

综上所述，对于同样的大小硅团簇，以[111]的作用晶向最为稳定；硅团簇与石墨烯复合结构稳定性将随着团簇的增大而增强，同时单键能减小；硅团簇的高度对结构稳定性影响甚微。

### 3.2 [111]晶向硅团簇和石墨烯复合模型储锂性能研究
#### 3.2.1 吸附能的计算

锂在复合结构中吸附相较与在单纯硅团簇中更加复杂。由上所述可知，当硅团簇的[111]晶向与石墨烯作用时最为稳定，因此我们考察了[111]晶向硅团簇与石墨烯的复合结构中不同位置嵌入锂原子的情况。如图 3(a)所示，共计算了 13 种储锂位置模型。其中编号为 H*的原子所在处为六角形中心（H），编号 T*的原子所在处为四面体中心（T），编号的数字从小到大对应着吸附的位置从低到高。

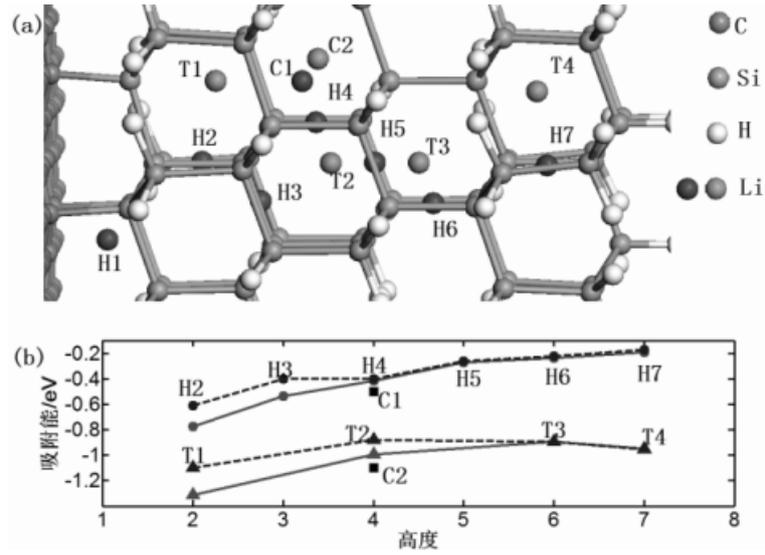

图 3 (a)复合结构中嵌锂的位置，(b)嵌锂吸附能(eV)

如图 3(b)给出了不同位置嵌锂吸附能 $E_b = E_{tot} - E_0 - \mu_{Li}$，与在单纯硅团簇中嵌锂吸附能进行了比较。式中 $E_{tot}$ 为嵌锂结构总能，$E_0$ 为不加锂的复合结构能量，$\mu_{Li}$ 为单个锂原子的自由能。四面体中心 T 位置的吸附能平均值约为—0.95 eV 比六边形中心 H 位置的吸附能(—0.4 eV)大，所以前者比后者更加容易的吸附锂，与 Zhang Q F 等的结果相符[11]。我们进一步选取了 6 个不同高度的 H 和 T 位置，研究锂的嵌入对于硅团簇的键长和键角产生的影响，结果见表 3。锂嵌入前后，键角最大变化为 2.1%，键长最大变化为 4.5%，表明单个锂的嵌入对于 Si 晶体结构有一定影响。对于六边形 H 位置，键长变化大于四面体 T 位置，而后者键角变化大于前者。这是因为 H 位置周围只有 6 个 Si 原子，其伸缩性较小，当嵌入锂时只能通过改变键长来容纳，而对于 T 位置周围有 10 个 Si 原子，形成了一个很大的笼状结构，其伸缩性很强，故在锂嵌入时对键角的改变较大。正是这种结构差异导致了 T 位置比 H 位置更加容易吸附锂。

表 3 锂嵌入前后键角和键长及变化

| 位置 | 键角（°） | | | 键长（Å） | | |
| --- | --- | --- | --- | --- | --- | --- |
| | 嵌锂前 | 嵌锂后 | 变化 | 嵌锂前 | 嵌锂后 | 变化 |
| H2 | 109.87 | 110.84 | 0.97 | 2.33 | 2.42 | 0.09 |
| H4 | 109.34 | 109.61 | 0.27 | 2.33 | 2.45 | 0.12 |
| H7 | 112.63 | 112.79 | 0.16 | 2.34 | 2.45 | 0.11 |
| T1 | 111.35 | 109.04 | 2.31 | 2.33 | 2.37 | 0.04 |
| T2 | 108.52 | 111.33 | 2.81 | 2.34 | 2.39 | 0.05 |
| T4 | 112.88 | 115.41 | 2.53 | 2.34 | 2.37 | 0.03 |

随着复合结构中嵌锂位置距石墨烯越近，吸附能变大。对于六边形中心 H 位情况，比较相同位置时对应的复合结构和单纯硅团簇的吸附能，在 H2 位置分别为－0.775 eV 和—0.609 eV，在 H3 位置分别为—0.537 eV 和—0.402 eV。可见由于硅团簇与石墨烯的相互作用，使得在靠近石墨烯的位置处吸附能变大，即石墨烯复合结构相对于硅团簇更有利于锂的吸附。这是因为在靠近石墨烯的位置，硅团簇为了与石墨烯形成较强的相互作用产生了一定的形变，这一形变使得硅的内部空间发生变化更加有利于锂的吸附。考察 H2 和 H4 位置在锂嵌入前后的 Muliken 电荷布居数，H2 位置周围的 Si 原子在锂嵌入后 Muliken 电荷布居数增加了 1.98%，H4 位置相应布居数增加了 1.67%。可见，较靠近石墨烯的 H2 位置从锂原子转移得到的电荷较距石墨烯较远的 H4 位置多，从而形成较强的吸附吸附作用。

在复合结构中 T4 位置处于硅团簇表面，形成的是表面吸附，其吸附能—0.947 eV 稍大于体内吸附时 T3 位置的—0.892 eV。我们也考察了相同高度距离表面不同距离的位置吸附锂的情况，如图 3(a)所示。C1 和 H4 为同一高度的六边形 H 位置，吸附能分别为—0.51 eV 和—0.418 eV。C2 和 T2 为同一高度四面体 T 位置，吸附能分别为—1.098 eV 和—0.992 eV。故对于相同高度和类型的位置，越接近表面吸附能越高，可见表面更加利于锂吸附。这是因为表面硅原子存在着悬链键，虽经过氢钝化活性仍较高，相较于体内硅原子更容易吸附锂。

H1 位置吸附能为—2.303 eV，接近于单纯石墨烯表面的锂吸附能。该位置距离石墨烯所在的平面为 1.86 Å，锂与石墨烯发生了较强的相互作用，提高了吸附能。考察 Muliken 电荷布居数，H1 位置锂原子电荷布居数为 1.493，H2 位置锂原子电荷布居数 1.694。相对于 H2 位置，较多的电荷从锂原子中转移出去。分析距 H1 位置最近的 C 原子的电荷布居数，发现锂嵌入后 C 原子的布居

数由 3.989 变为 4.103，说明 C 原子和锂原子之间发生了一定的电荷转移，提高了吸附能。我们进一步研究了 T1 位置的锂与距离其最近的碳和硅的 PDOS 图，如图 4 所示。锂原子的 2s 轨道主要分布在—0.5 eV 以上，C 原子的 2p 轨道、Si 原子的 3p 轨道从—5 eV 到 3 eV 都有分布。锂原子的 2s 轨道与 C 原子的 2p 轨道、Si 原子的 3p 有很大程度的交叠，可见 T1 位置的锂原子和近邻的硅原子、碳原子发生较为强烈的相互作用。

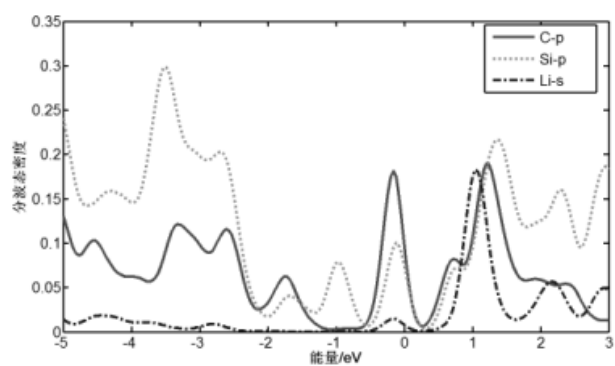

图 4 T1 位置最紧邻的 Li、C 和 Si 原子的 PDOS 图

### 3.2.2 锂嵌入产生的结构形变

与普通的硅团簇相比，通过计算在同一个高度同时嵌入两个锂原子后的形变，我们进一步研究了复合结构用于锂电池负极时的循环性能。如图 5(a)所示的结构中，分别在两个相邻的四面体 T 位置各嵌入一个锂原子。由于石墨烯使得硅团簇的分布变得分散，锂嵌入后的晶格张量很小，用其来衡量锂嵌入后硅团簇的形变将存在较大的误差。因此，为了更好地衡量结构的形变，我们计算了与锂原子紧邻的四个 Si 原子组成的四面体在锂的嵌入后体积的变化，如图 5(b)所示。

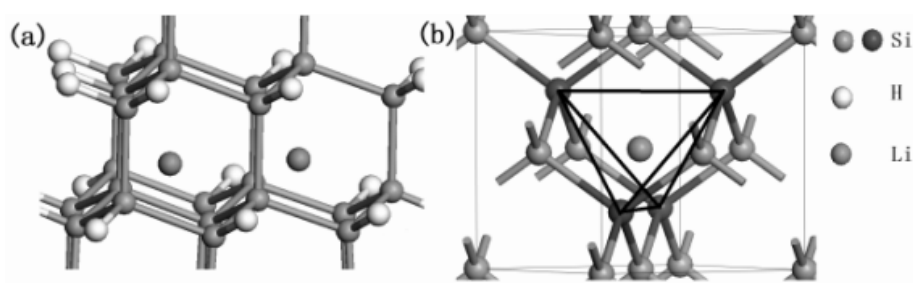

图 5(a)锂嵌入的位置，(b)与锂紧邻的硅组成的四面体示意图

对于硅团簇 T1 和 T2 位置四面体体积变化分别为 21.57%和 12.24%，复合结构分别为 15.97%和 12.17%。其体积变化量均大于 10%，说明锂的嵌入对于结构有着一定破坏。因为计算的是锂原子周围的 Si 原子组成的四面体产生的形变，此形变量大致相当于整个体系充满锂时的体积形变量。对于硅团簇，相对于 T2 位置，T1 位置处于表面形变较大。对于复合结构，在距离石墨较远的 T2 位置形变量为 12.17%，与硅团簇的 12.24%差别很小。而靠近石墨烯的 T1 位置，硅团簇和石墨烯形成较强的键，对团簇起到固定作用，相比于硅团簇的形变量 21.57%，复合结构降到了 15.97%。可见石墨烯的引入较好地抑制了锂的嵌入而产生的形变，有利于提高结构的刚性。

## 4. 结论

采用基于密度泛函理论的第一性原理计算，系统研究了不同高度、不同尺寸、不同接触晶向、不同成键浓度的硅团簇和石墨烯的复合结构稳定性，并研究其储锂性能。分析结果表明，硅团簇和石墨烯之间可以形成较强的 Si-C 键，从而形成稳定的结构。随着硅团簇尺寸的增大将产生更多不匹配原子，破坏结构的有序性，从而使单根的 Si-C 键形成能降低。接触面为[111]晶向的硅团簇与石墨烯形成的复合结构最为稳定。由于影响结合能的主要因素为 Si-C 之间的相对位置和成键情况，因而硅团簇的高度对形成能的影响并不显著。对复合结构的嵌锂吸附能计算得出，硅团簇中储锂位置靠近石墨烯时吸附能会大于远离石墨烯的储锂位置，更加有利于锂的存储。而界面处锂吸附时与 Si 和 C 相互作用较为明显。另外，引入石墨烯的结构使得在锂的嵌入过程中产生的结构破坏有明显的降低。所以，硅纳米团簇与石墨烯复合结构被证明是适合用于锂电池的储锂材料。

## 参考文献

# First-principles study on the Li storage performance of silicon clusters and graphene composite structure*

Wu Jiangbin, Qian Yao, Guo Xiaojie, Cui Xianhui, Miao Ling[†], Jiang Jianjun

*(Electronic of science and technology of Huazhong University of Science and Technology,430074, China)*

Abstract

This paper focuses on the performance of the storage of Li and the stability of the hybrid structure of different lattice planes of the silicon clusters and graphene by the first-principle theory. In this paper, we calculate the binding energy, adsorption energy and PDOS of the hybrid structure of the different height and size of the silicon clusters and graphene. We figure out that there can form strong Si-C bonds between the silicon cluster and graphene. Especially, the hybrid structure of the silicon clusters with plane (111) and graphene performs best with the highest formation energy and the outstanding stability. According to the calculation of Li absorption energy, we conclude that the location of the silicon cluster near the graphene has higher possibility and higher absorption energy of the Li storage, with the reason that the charge transfer between the lithium and the carbon and the silicon. Because of the graphene used, the deformation of the interface of the silicon clusters can be obviously reduced during the absorption of Li, which brings about a good future for the hybrid structure using for the battery anode materials.

**Key words:** silicon cluster, graphene, first-principles, lithium ion battery

**PACC:** 6140M, 6855, 7115A, 8640H

**PACS:** 36.40.Cg, 63.20.dk, 71.15.Mb, 84.60.Ve

* Project supported by the National Natural Science Foundation of China (Grant No. 50771047).
[†] Email: miaoling@hust.edu.cn, Tel: 027-87544472

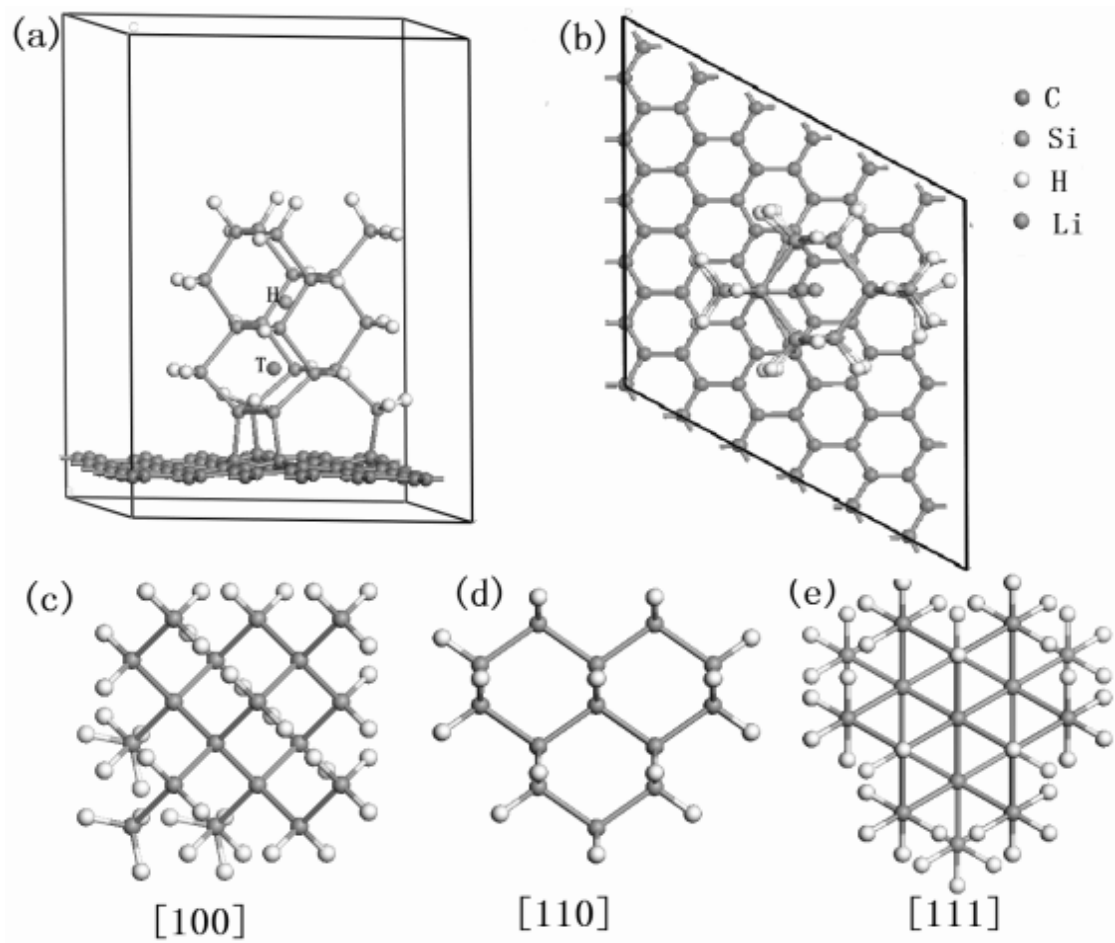

图 1 硅纳米团簇石墨烯复合模型(a)侧视图(b)顶视图。(c)(d)(e)不同硅团簇晶向

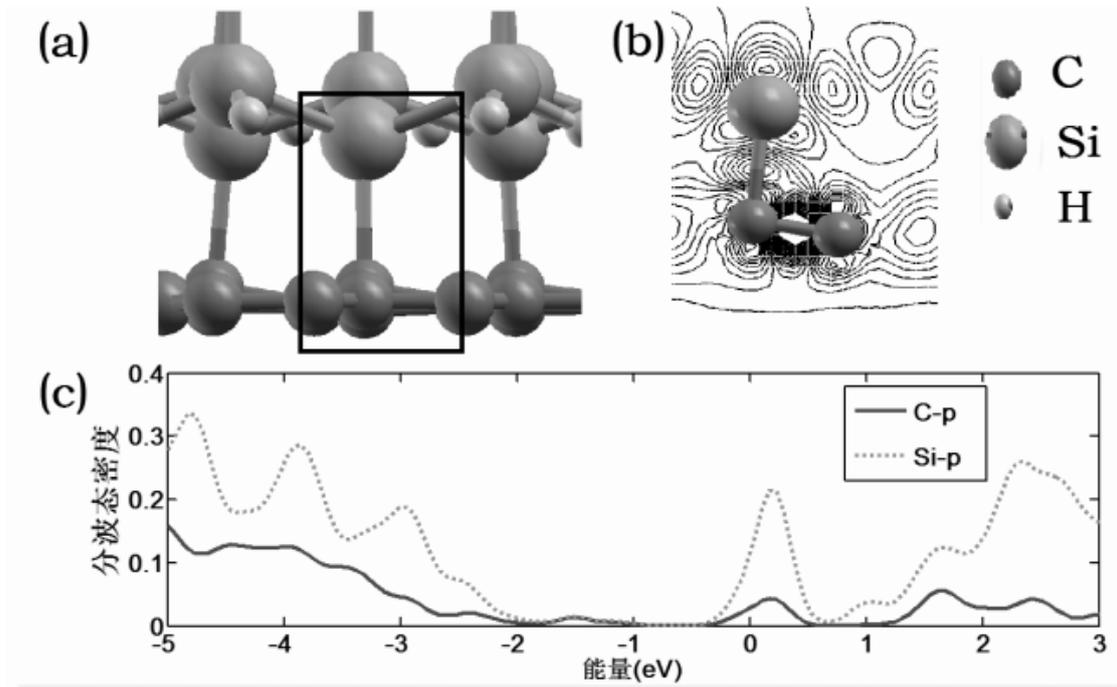

图 2 (a)硅纳米团簇与石墨烯相互作用情况，(b)PDOS 图

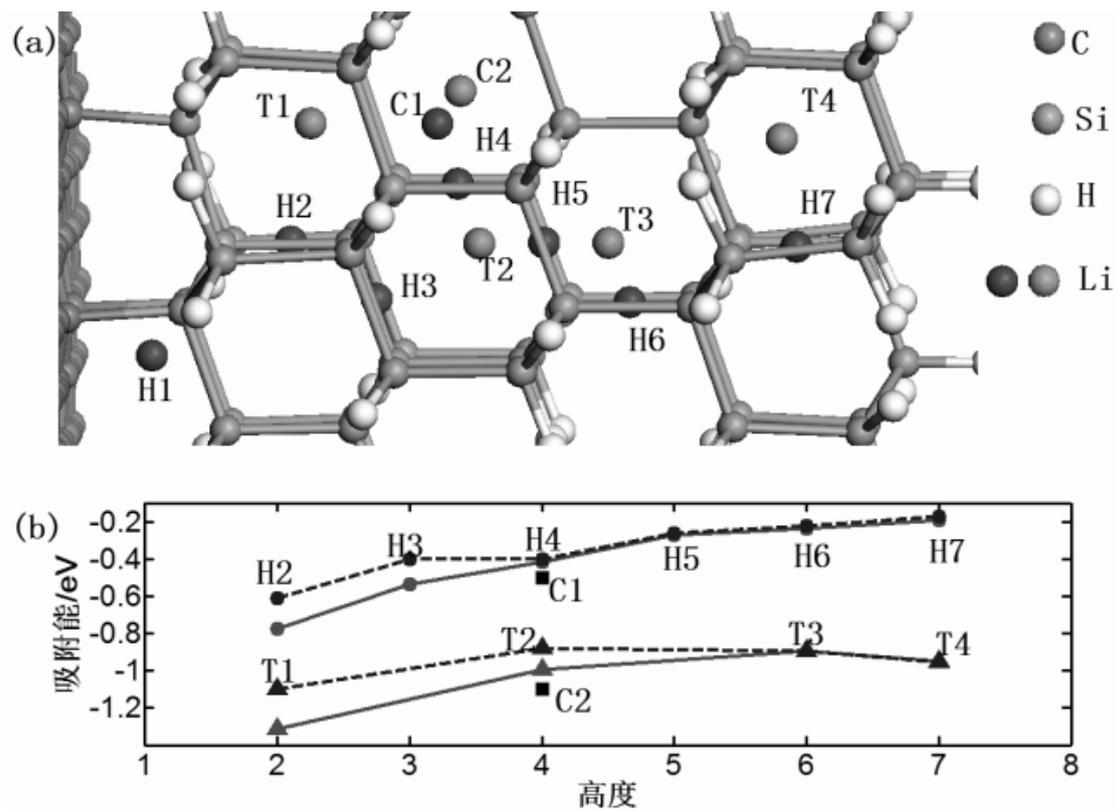

图 3 (a)复合结构中嵌锂的位置，(b)嵌锂吸附能(eV)

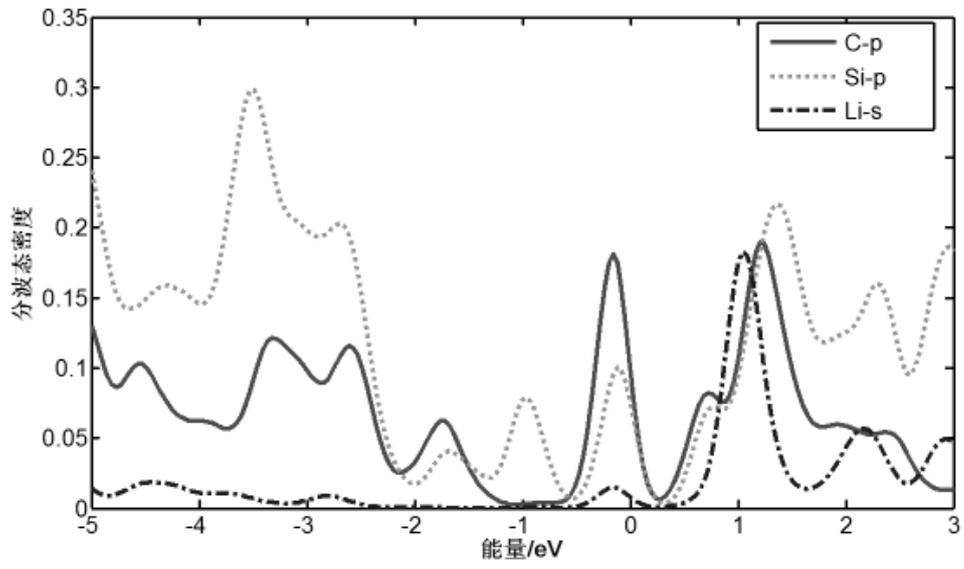

图 4 T1 位置最紧邻的 Li、C 和 Si 原子的 PDOS 图

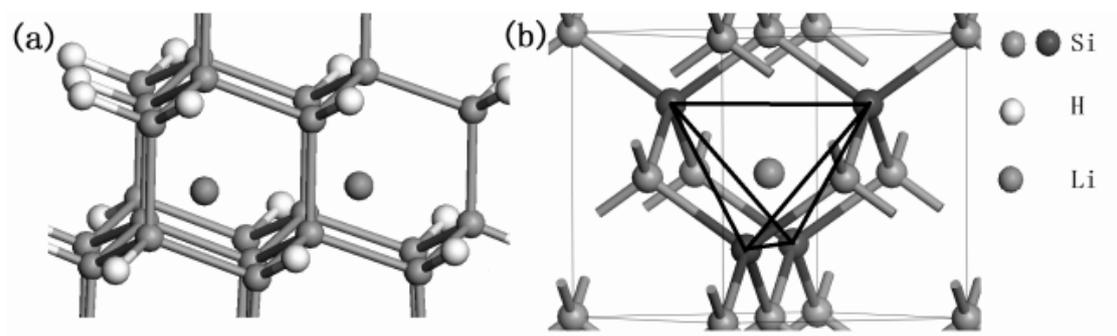

图 5 (a)锂嵌入的位置，(b)与锂紧邻的硅组成的四面体示意图